Electron-phonon coupling superconductivity in 2D orthorhombic $MB_6$ (M=Mg, Ca) and hexagonal $MB_6$ (M=Mg, Ca, Sc, Ti, Sr, Y)

Tao Bo,[1,2] Peng-Fei Liu,[1] Luo Yan,[1] and Bao-Tian Wang[1,3,*]

[1]Institute of High Energy Physics, Chinese Academy of Sciences (CAS), Beijing 100049, China

[2]Songshan Lake Materials Laboratory, Dongguan, Guangdong, 523808, china

[3]Collaborative Innovation Center of Extreme Optics, Shanxi University, Taiyuan, Shanxi 030006, China

*Corresponding author: wangbt@ihep.ac.cn

**ABSTRACT:**

Combining crystal structure search and first-principles calculations, we report a series of two-dimensional (2D) metal borides including orthorhombic (ort-) $MB_6$ (M=Mg, Ca) and hexagonal (hex-) $MB_6$ (M=Mg, Ca, Sc, Ti, Sr, Y). Then, we investigate their geometrical structures, bonding properties, electronic structures, mechanical properties, phonon dispersions, thermal stability, dynamic stability, electron-phonon coupling (EPC), superconducting properties and so on. Our *ab initio* molecular dynamics simulation results show that these $MB_6$ can maintain their original configurations up to 700/1000 K, indicating their excellent thermal stability. All their elastic constants satisfy the Born mechanically stable criteria and no visible imaginary frequencies are observed in their phonon dispersions. The EPC results show that these 2D $MB_6$ are all intrinsic phonon-mediated superconductors with the superconducting transition temperature ($T_c$) in the range of 2.2-21.3 K. Among them, the highest $T_c$ (21.3 K) appears in hex-$CaB_6$, whose EPC constant ($\lambda$) is 0.94. By applying tensile/compressive strains on ort-/hex-$CaB_6$, we find that the compressive strain can obviously soften the acoustic phonon branch and enhance the EPC as well as $T_c$. The $T_c$ of the hex-$CaB_6$ can be increased from 21.3 K to 28 K under compressive strain of 3%. These findings enrich the database of 2D superconductors and should stimulate experimental synthesizing and characterizing of 2D superconducting metal borides.

## I. INTRODUCTION

Superconducting materials have excellent physical and chemical properties, and may be widely used in electric power, electronics, medical, transportation, high-energy physics and other fields. In addition to the hot topics of copper oxide [1] and iron-based superconductors [2], conventional electron-phonon superconductors are still attractive [3]. Among reported conventional phonon-mediated superconductors, the anisotropic two-gap superconductor $MgB_2$ ever holds the record of the highest superconducting transition temperature ($T_c$) of 39 K [4]. This discovery has stimulated many efforts in exploring the microscopic mechanisms [5-7] and also in searching new conventional superconductors [8, 9]. The predicted superconductors of hydrogen sulfide [10] and $LaH_{10}$ [11], based upon the Bardeen-Cooper-Schrieffer (BCS) theory [12], have been experimentally verified [13-15]. Their $T_c$ of 203 and 250 K have greatly overcome the $T_c$ of $MgB_2$.

Reducing dimensionality of materials can modulate their properties and would also produce new structures. With the development of the low-dimensional nanostructure technology, many two-dimensional (2D) superconductors have been successfully synthesized [16, 17], such as α-$Mo_2C$ [18], Li-decorated monolayer graphene [19], $NbSe_2$ [20], potassium-intercalated $T_d$-$WTe_2$ [21], stanene [22], and magic twisted graphene [23]. Fascinating of Kosterlitz–Thouless–Berezinskii transition, electron quantum confinement effects, and charge density wave (CDW) have been observed in these 2D superconductors. Thus, it is of great interesting to find new 2D superconductors and investigate the underlying superconducting mechanism for not only the basic scientific studies but also the application of superconducting electronic devices in high and new technology. Based upon density functional theory (DFT) and using the BCS theory, many potential 2D superconductors have been predicted, such as the borophene [24-27], aluminum-deposited graphene [28], carrier-doped or strained graphene [29, 30], $\beta_0$-PC [31], $MgB_x$ [32], $Mo_2B_2$ [33], $W_2B_2$ [34], $AlB_6$ [35]. Interestingly, competition or coexistence of the CDW and superconducting order in 2D $TiSe_2$ [36] and $NbSe_2$ [37], hydrogenation superconducting monolayer $MgB_2$ [38], Na-intercalated superconducting $MoX_2$ (X=S, Se) bilayers [39], Ca-intercalated $\beta$-Sb bilayer superconductor [40], electron/hole-doped superconducting $PtSe_2$ [41], and Br-functionalized superconducting monolayer $Mo_2C$ [42] have been studied.

The superconductivity with high $T_c$ may exist in materials composed of light atoms [43]. Boron has a very small atomic mass, which may enable superconducting behaviors in its elemental or alloying forms. Experimentally, a high $T_c$ of 36 K was observed in boron-doped Q-carbon [8]. In bulk $YB_6$, a $T_c$ of ~7 K was reported through measuring its specific heat, resistivity, and magnetic susceptibility [44]. Till now, superconducting properties have never been experimentally observed in 2D boron and 2D borides. The difficulty may in the synthesizing and also in measuring. However, in recent years, many borophenes and 2D metal borides have been predicted to be superconductors, such as $\beta_{12}$ borophene [24-26], $\chi_3$ borophene [25], $B_xMgB_x$ (x=2-5) borides [32] and $AlB_6$ nanosheet [35]. The $T_c$ of $\beta_{12}$ borophene is greatly suppressed by biaxial tensile strain induced by Ag(111) substrate, which explains to some extent the difficulty in experimental measurement [26]. The newly predicted orthogonal $AlB_6$ nanosheet not only possesses triple Dirac cones, Dirac-like fermions, and node-loop features, but also has superconductivity with $T_c$ = 4.7 K [35].

Recently, we also predicted some novel 2D superconductors of $Mo_2B_2$ [33], $W_2B_2$ [34], and $XB_6$ (X=Ga, In) [45] based on the crystal structure prediction method and first-principles calculations. A certain tensile/compressive strain can regulate the EPC as well as $T_c$ of these materials. In the present work, we first report a series of 2D $MB_6$ structures including orthogonal (ort-) $MgB_6$, $CaB_6$, and hexagonal (hex-) $MgB_6$, $CaB_6$, $ScB_6$, $TiB_6$, $SrB_6$, and $YB_6$ through structural searching by using the CALYPSO code. Then, we investigate their crystal structures, electronic structures, phonon dispersions, mechanical properties, EPC, and superconducting properties by performing first-principles calculations. We find that these 2D metal hexaborides are all phonon-mediated superconductors with $T_c$ in range of 2.2-21.3 K. The underlying mechanism of their superconductivity is carefully analyzed. In addition, the effect of tensile/compressive strain on the EPC constant and superconducting properties is also studied.

## II. COMPUTATIONAL DETAILS

We search for 2D $MB_6$ (M= Mg, Ca, Sc, Ti, Sr, Y) using the particle swarm optimization (PSO) scheme as implemented in the CALYPSO code [46-48]. In searching, both planar and buckled structures including one, two, and three layers are considered. For each $MB_6$ system, one, two, and four formula units per simulation cell are calculated. The number of generations are set to be 30 and each generation contains 30 structures. After structural searching, thousands of 2D $MB_6$ structures are generated. The subsequent structural optimization, energy calculations, and *ab initio* molecular dynamics (AIMD) simulations are carried out by utilizing the density functional theory (DFT) method as implemented in the Vienna Ab initio Simulation Package (VASP) [49]. The Perdew-Burke-Ernzerhof generalized gradient approximation (PBE-GGA) [50] is employed and the electron-ion interaction is described by using the projector augmented wave (PAW) method [51]. The plane-wave cutoff energy is set as 600 eV and the Brillouin zone (BZ) integration is performed using a $11 \times 11 \times 1$ Monkhorst-Pack (MP) mesh. All the geometries are relaxed until all forces on the atoms are smaller than 0.005 eVÅ$^{-1}$. A vacuum separation is set to more than 20 Å to prevent any interactions between two neighboring monolayers. The van der Waals (vdW) correction is included by using the optB86b-vdW exchange functional [52] since it can properly treat the long-range dispersive interactions. The AIMD simulations in the *NVT* ensemble are employed at 300, 700, and 1000 K to investigate the thermal stability of these 2D $MB_6$ monolayers.

They are performed by using $3 \times 3$ supercells with a time step of 1 fs and total simulation of 5 ps.

The calculations of band structure, phonon dispersion, EPC, and superconducting property are performed at the DFT level, employing the norm-conserving pseudopotentials [53] as implemented in the QUANTUM-ESPRESSO (QE) package [54]. The VASP-optimized structures are re-optimized within QE. The plane-waves kinetic-energy cutoff and the energy cutoff for charge-density are set as 80 and 320 Ry, respectively. The Methfessel-Paxton smearing width of 0.02 Ry is used. The BZ *k*-point grid of 32×32×1 is adopted for the self-consistent electron density calculations. The dynamic matrix and EPC matrix elements are calculated on a $8 \times 8 \times 1$ *q*-point mesh for both *ort*- and *hex*-$MB_6$ monolayers. The phonon properties and EPC are calculated within the density-functional perturbation theory [55] and Eliashberg theory [56].

The EPC $\lambda_{qv}$ is calculated according to the Migdal-Eliashberg theory [57] by

$$\lambda_{qv} = \frac{\gamma_{qv}}{\pi h N(E_F)\,\omega_{qv}^2}, \tag{1}$$

where $\gamma_{qv}$ is the phonon linewidth, $\omega_{qv}$ is the phonon frequency, and $N(E_F)$ is the electronic density of state at the Fermi level. The $\gamma_{qv}$ can be estimated by

$$\gamma_{qv} = \frac{2\pi\omega_{qv}}{\Omega_{BZ}} \sum_{k,n,m} \left| g_{kn,k+qm}^{v} \right|^2 \delta\left(\epsilon_{kn} - \epsilon_F\right) \delta\left(\epsilon_{k+qm} - \epsilon_F\right), \tag{2}$$

where $\Omega_{BZ}$ is the volume of BZ, $\epsilon_{kn}$ and $\epsilon_{k+qm}$ indicate the Kohn-Sham energy, and $g_{kn,k+qm}^{v}$ represents the EPC matrix element. The Eliashberg spectral function $\alpha^2F(\omega)$ can be estimated by

$$\alpha^2 F(\omega) = \frac{1}{2\pi N(E_F)} \sum_{qv} \frac{\gamma_{qv}}{\omega_{qv}} \delta\left(\omega - \omega_{qv}\right). \tag{3}$$

Then, the EPC constant $\lambda(\omega)$ and the logarithmic average frequency $\omega_{\log}(\omega)$ can be determined by

$$\lambda(\omega) = 2\int_0^{\omega} \frac{\alpha^2 F(\omega)}{\omega} d\omega \tag{4}$$

and

$$\omega_{\log}(\omega) = \exp\left[\frac{2}{\lambda}\int_0^{\infty} \frac{d\omega}{\omega}\alpha^2 F(\omega)\log\omega\right] . \tag{5}$$

After obtaining the total λ and $\omega_{\log}$, the superconducting transition temperature $T_c$ can be calculated by

$$T_c = \frac{\omega_{\log}}{1.2}\exp\left[-\frac{1.04(1+\lambda)}{\lambda-\mu^*(1+0.62\lambda)}\right]. \tag{6}$$

In our following calculations, a typical value of the effective screened Coulomb repulsion constant $\mu^* = 0.1$ is used [24, 58].

## III. RESULTS AND DISCUSSION

### A. Geometrical structures and bonding nature of the ort- and hex-$MB_6$

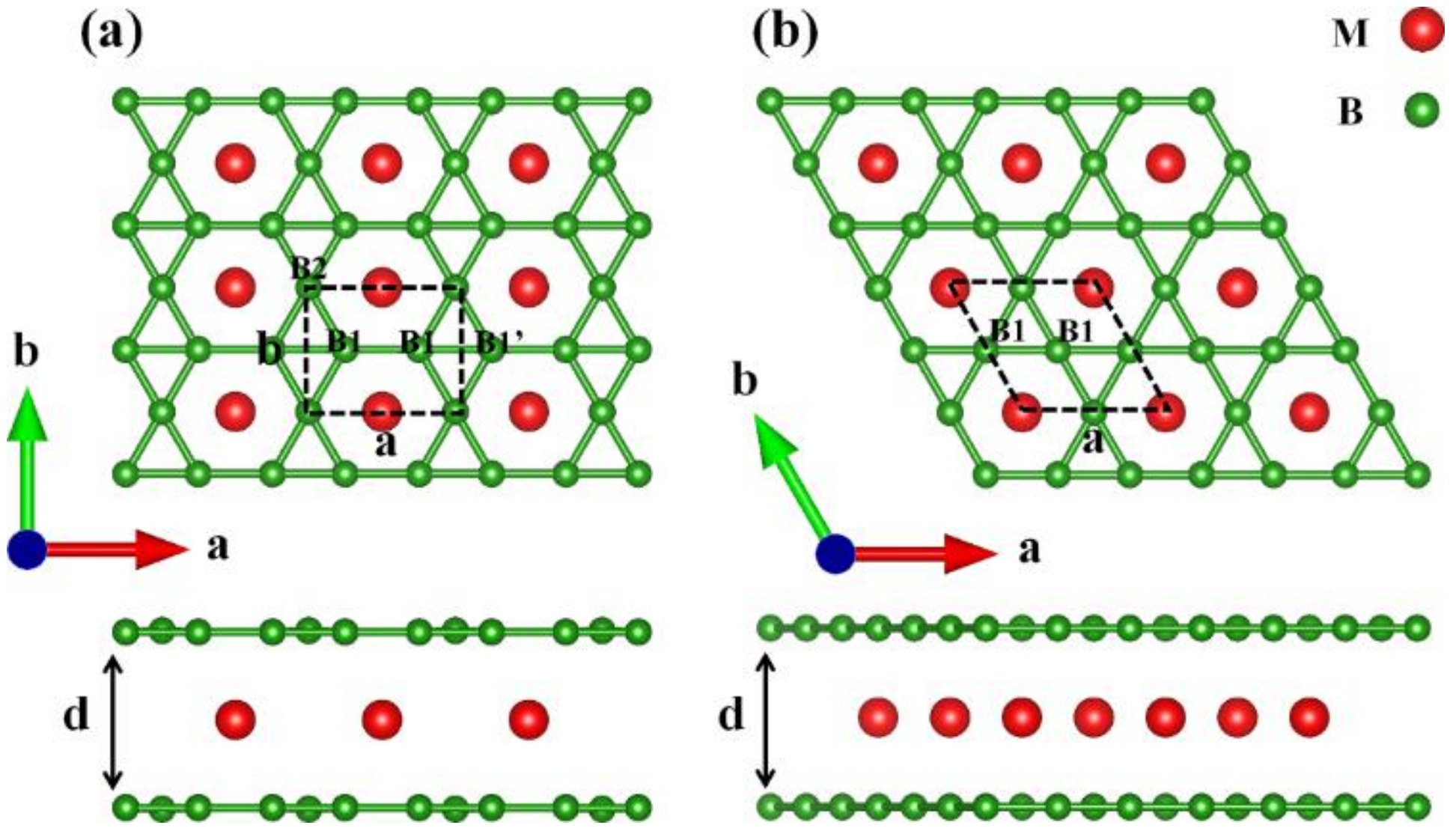

**FIG. 1.** Top and side views of the (a) ort-$MB_6$ and (b) hex-$MB_6$ monolayers.

We perform comprehensive structural search using the CALYPSO code to find various configurations of 2D $MB_6$. The structural search results show that the orthogonal configuration [see Fig. 1(a)] is the global minimum structure for all these 2D $MB_6$. The energies of the ort-$MB_6$ in sequence of M = Mg, Ca, Sc, Ti, Sr, Y are 1.060, 0.964, 0.223, 0.006, 1.029, and 0.193 eV per unit cell lower than those of their corresponding hexagonal phase, respectively. In one recent theoretical work [32], the energy of the ort-$MgB_6$ was also reported lower by 1.02 eV per unit cell than that of the hex-$MgB_6$. The global minimum structure of the 2D $AlB_6$ [35] is also the orthogonal one. However, the phonon spectra results calculated by VASP as well as QE show that the orthogonal structure of $ScB_6$, $TiB_6$, $SrB_6$, and $YB_6$ is dynamically unstable. As known, the global minimum energy is not the exclusive condition for experimental synthesis of materials. Under suitable conditions, it is possible to synthesize some metastable crystals with higher energy [59]. On the basis of aforementioned discussions, in the following, we will focus on the (meta)stable structures including the orthogonal $MB_6$ (M = Mg, Ca) and hexagonal $MB_6$ (M = Mg, Ca, Sc, Ti, Sr, Y) for further analysis.

The ort-$MB_6$ [see Fig. 1(a)] and hex-$MB_6$ [see Fig. 1(b)], consisting of one layer of metal atoms and two layers of B atoms, belong to the orthogonal *Pmmm* (No. 47) and the hexagonal *P6/mmm* (No. 191) space group, respectively. For the ort- and hex-$MB_6$, their B layers are borophenes with hexagonal vacancy density ($\eta$) of 1/4, named $\delta_4$ [60] and kagome [61], respectively. The metal atoms, which keep the two B layers from fusing together, are sandwiched by top and bottom B layers. Meanwhile, the metal atoms are anchored and stabilized by the two B layers, leading to the high stability of these 2D materials. As shown in Table I, the optimized lattice parameters are $a$=3.46, $b$=2.91 Å for the ort-$MgB_6$ and $a$=3.48, $b$=2.94 Å for the ort-$CaB_6$, which are comparable to those of $AlB_6$ ($a$=3.41, $b$=2.91 Å) [35]. However, the thickness of the ort-$CaB_6$ (4.24 Å) is much larger than that of the ort-$MgB_6$ (3.55 Å). Same phenomenon also occurs in the hex-$MB_6$. The lattice constants of these hexagonal structures vary slightly, ranging from $a$ = $b$ = 3.40 to 3.44 Å, while the thicknesses vary considerably, ranging from 3.33 to 4.65 Å. The lattice constant $a$ of the hex-$MgB_6$ (3.40 Å) in our work is the same as that of Xie et al.'s work [61]. We find that different metal layers have little effect on the lattice constants $a$ and $b$ but have relatively large effect on the thickness of both 2D ort- and hex-$MB_6$. The strong bonding among B atoms makes the lattice constants almost unchanged and the variety of bond lengths between metal atoms and B atoms leads to the visible change of the cell thicknesses. For the ort-$MB_6$, there are two symmetrically distinct atom types for B [see Fig. 1(a), labeled as B1 and B2]. As shown in Table I, the Mg-B1/Ca-B1 and Mg-B2/Ca-B2 distances are 2.50/2.70 and 2.48/2.74 Å, respectively, which are a little larger than those of the $AlB_6$ (2.409 and 2.262 Å, respectively) [35]. The B1-B2 bond length of the ort-$MgB_6$/$CaB_6$ is both 1.70 Å, almost the same as that of the $AlB_6$ (1.708 Å). There are two types of B1-B1 bonds in ort-$MB_6$: the bond along one side of the hexagonal holes (B1-B1) and the bond along the diagonal of the rhombus (B1-B1'). The B1-B1 and B1-B1' distances for the ort-$MgB_6$/$CaB_6$ are 1.73/1.76 and 1.73/1.72 Å, respectively, closing to those of the $AlB_6$ (1.688 and 1.723 Å, respectively). For the hex-$MB_6$, there is just one symmetrically distinct atom type for B [see Fig. 1(b), labeled as B1]. The B1-B1 distances for these six hex-$MB_6$ monolayers are almost the same, in range of 1.70-1.72 Å. However, the M-B1 distances vary considerably, ranging from 2.38 to 2.89 Å.

To get more information about the bonding, we calculate the line charge density distribution along the nearest B-B and M-B bonds and perform the electron localization function (ELF) [62] and Bader charge analysis [63]. The ELF (Fig. S1) pictures as well as the values of Bader charge ($Q_B$) and charge density at the corresponding bond point ($CD_b$) (Table S1) are presented in the supporting information (SI). For the ort-/hex-$MgB_6$, hex-$ScB_6$, and hex-$TiB_6$, there are mainly covalent bonds between B atoms and ionic bonds between metal atoms and the neighboring B atoms. However, for the ort-/hex-$CaB_6$, hex-$SrB_6$, and hex-$YB_6$, in addition to the above chemical bonds, there are also metallic bonds between M atoms. One can see the detailed discussions in the SI.

TABLE I. Lattice parameters (Å), bond lengths (Å), and cohesive energies (eV) of the ort- and hex-$MB_6$.

| | a | b | d | B1-B1/B1-B1' | B1-B2 | M-B1 | M-B2 | M-M | Cohesive Energy |
|---|---|---|---|---|---|---|---|---|---|
| Ort-$MgB_6$ | 3.46 | 2.91 | 3.55 | 1.73/1.73 | 1.70 | 2.50 | 2.48 | 2.91/3.46 | 5.42 |
| Ort-$CaB_6$ | 3.48 | 2.94 | 4.24 | 1.76/1.72 | 1.70 | 2.70 | 2.74 | 2.94/3.48 | 5.55 |
| Hex-$MgB_6$ | 3.40 | | 3.63 | 1.70 | | 2.49 | | 3.40 | 5.27 |
| Hex-$CaB_6$ | 3.42 | | 4.26 | 1.71 | | 2.73 | | 3.42 | 5.41 |
| Hex-$ScB_6$ | 3.41 | | 3.73 | 1.70 | | 2.53 | | 3.41 | 6.00 |
| Hex-$TiB_6$ | 3.40 | | 3.33 | 1.70 | | 2.38 | | 3.40 | 6.32 |
| Hex-$SrB_6$ | 3.44 | | 4.65 | 1.72 | | 2.89 | | 3.44 | 5.28 |
| Hex-$YB_6$ | 3.44 | | 4.15 | 1.72 | | 2.69 | | 3.44 | 5.93 |

### B. Stability and mechanical properties of the ort- and hex-$MB_6$

It is important to verify the stability of these $MB_6$ monolayers, which is helpful for judging whether they can be synthesized experimentally. Cohesive energy is a widely accepted parameter used to evaluate the stability of materials. We calculate the cohesive energy $E_{coh}$ using the following formulas:

$$E_{coh} = \frac{E_{M} + 6E_{B} - E_{\mathbf{MB_6}}}{7} \quad (7)$$

where $E_M$ and $E_B$ are the total energies of the isolated metal and B atoms, respectively, $E_{\mathbf{MB_6}}$ is the total energy of the 2D $MB_6$. As shown in Table I, the calculated cohesive energies are 5.42, 5.55, 5.27, 5.41, 6.00, 6.32, 5.28, and 5.93 eV/atom for the ort-$MgB_6$, ort-$CaB_6$, hex-$MgB_6$, hex-$CaB_6$, hex-$ScB_6$, hex-$TiB_6$, hex-$SrB_6$, and hex-$YB_6$, respectively. These values are larger than those of the previously computationally predicted and later experimentally fabricated 2D materials, such as $Cu_2Si$ (3.46 eV/atom) [64, 65], $Ni_2Si$ (4.80 eV/atom) [66], and $Cu_2Ge$ (3.17 eV/atom) [67]. The relatively large cohesive energies indicate that these $MB_6$ monolayers can be synthesized under appropriate experimental conditions. We also calculate the phonon dispersions along the high-symmetry lines as well as the phonon density of states (PhDOS) to evaluate the dynamic stability of these $MB_6$ monolayers. For the ort- and hex-$MgB_6$, ort- and hex-$CaB_6$, hex-$ScB_6$, and hex-$TiB_6$, the fact that no imaginary modes in the BZ confirms the lattice dynamic stability of these six $MB_6$ monolayers. A detailed discussion of the phonon dispersions for all the eight $MB_6$ monolayers will be supplied later. In addition, thermal stability is very important for the practical application of these 2D $MB_6$ in nano-electronic devices or electrode films. In order to investigate the thermal stability of these 2D $MB_6$, AIMD simulations are performed with total simulation time of 5 ps at 300, 700, and 1000 K. We present in Fig. S2, S3, and S4 the variation of total energies with time for the $MB_6$ monolayers. The snapshots after 5 ps for these simulated $MB_6$ are shown in Fig. S5 and S6. We find that the total energy of each $MB_6$ monolayer fluctuates around a constant value at 300 and 700/1000 K. After simulating for 5 ps, the framework of each $MB_6$ monolayer well maintains its original configuration at both 300 and 700/1000 K. The above results indicate the excellent thermal stability of these $MB_6$ monolayers.

Mechanical properties are also very important for the practical application of 2D materials. In order to accurately understand the mechanical properties of these $MB_6$ monolayers, we calculate their elastic constants [68]. Generally speaking, a mechanically stable 2D material needs to satisfy the Born criteria [69], $C_{11}C_{22} - C_{12}^2 > 0$ and $C_{66} > 0$. For hexagonal symmetry, $C_{11} = C_{22}$. Thus, the stability criteria is $|C_{11}| > |C_{12}|$ and $C_{66} > 0$. The calculated elastic constants of these eight $MB_6$ monolayers are listed in Table II. We find that the elastic constants of these $MB_6$ monolayers all meet the Born criteria, indicating that they are mechanically stable. Based on the elastic constants, the Young's moduli along the *x* and *y* directions of these materials are calculated according to $Y_x = \frac{(C_{11}C_{22} - C_{12}C_{21})}{C_{22}}$ and $Y_y = \frac{(C_{11}C_{22} - C_{12}C_{21})}{C_{11}}$. For the ort-$MgB_6$ and ort-$CaB_6$, the values of $Y_y$ are much larger than that of the graphene (340 N/m) [70], indicating that these two monolayers have excellent mechanical properties along the *y* direction. For the hexagonal $MB_6$ monolayers, the values of $Y_x/Y_y$ differ a little from that of the graphene (340 N/m), also exhibiting excellent mechanical properties. We also calculate the Poisson's ratio by $v_x = \frac{C_{21}}{C_{22}}$ and $v_y = \frac{C_{12}}{C_{11}}$ based on the calculated elastic constants. For the ort-$MgB_6$ and ort-$CaB_6$, the $v_x$ and $v_y$ values (0.060-0.094) are much smaller than that of the graphene ($v$=0.173) [71], indicating that the compression in $y/x$ direction is very small when stretching in $x/y$ direction. For the hexagonal $MB_6$ monolayers except hex-$TiB_6$, the $v_x/v_y$ values (0.190-0.210) are similar to that of graphene ($v$=0.173). However, for the hex-$TiB_6$, the $v_x/v_y$ values ($v$=0.296) are much larger than that of graphene ($v$=0.173), indicating large compression in $y/x$ direction when stretching in $x/y$ direction.

TABLE II. Calculated elastic constants $C_{ij}$ (N $m^{-1}$), Young's moduli (N $m^{-1}$), and Poisson's ratio of the ort- and hex-$MB_6$.

| | $C_{11}$ | $C_{22}$ | $C_{12}$ | $C_{66}$ | $Y_x$ | $Y_y$ | $v_x$ | $v_y$ |
|---|---|---|---|---|---|---|---|---|
| Ort-$MgB_6$ | 361.1 | 404.8 | 34.0 | 110.1 | 358.2 | 401.6 | 0.084 | 0.094 |
| Ort-$CaB_6$ | 321.2 | 395.6 | 23.8 | 129.3 | 319.8 | 393.8 | 0.060 | 0.074 |
| Hex-$MgB_6$ | 344.7 | | 67.9 | 145.8 | 331.3 | 331.3 | 0.197 | 0.197 |
| Hex-$CaB_6$ | 340.1 | | 66.0 | 140.3 | 327.3 | 327.3 | 0.194 | 0.194 |
| Hex-$ScB_6$ | 357.5 | | 68.1 | 142.7 | 344.5 | 344.5 | 0.190 | 0.190 |
| Hex-$TiB_6$ | 315.5 | | 93.3 | 107.5 | 287.9 | 287.9 | 0.296 | 0.296 |
| Hex-$SrB_6$ | 310.9 | | 65.3 | 122.6 | 297.2 | 297.2 | 0.210 | 0.210 |
| Hex-$YB_6$ | 307.3 | | 61.5 | 120.5 | 295.0 | 295.0 | 0.200 | 0.200 |

## C. Electronic properties

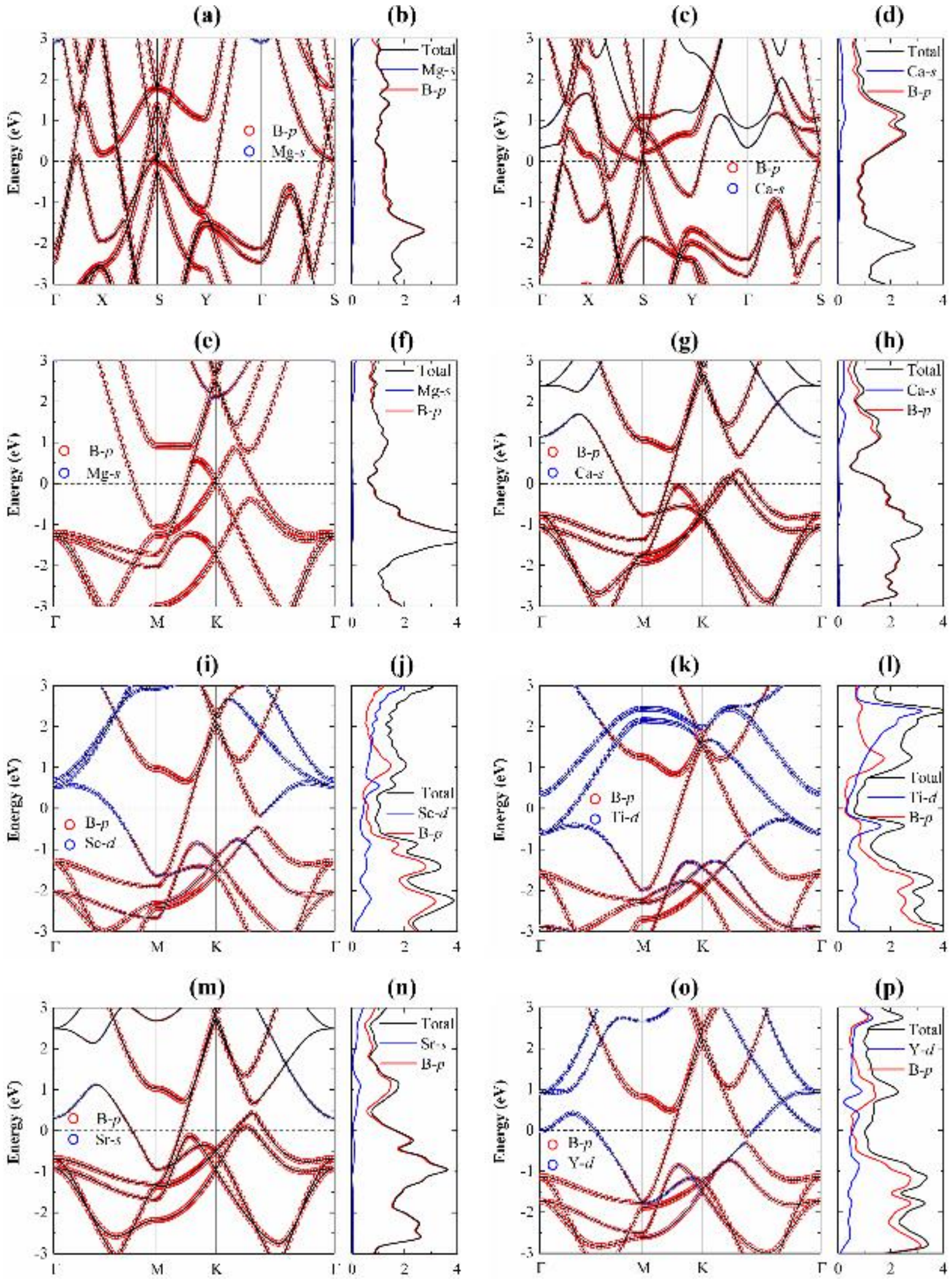

FIG. 2. Orbital-resolved band structures and corresponding total and local DOSs of (a-b) ort-$MgB_6$, (c-d) ort-$CaB_6$, (e-f) hex-$MgB_6$, (g-h) hex-$CaB_6$, (i-j) hex-$ScB_6$, (k-l) hex-$TiB_6$, (m-n) hex-$SrB_6$, and (o-p) hex-$YB_6$.

For all these $MB_6$ monolayers, we show the orbital-resolved band structures and electronic density of states (DOS) in Fig. 2. The corresponding Fermi surfaces (FS) are shown in Fig. 3. We find that all these materials exhibit metallic properties because there are many bands crossing the Fermi energy level. For ort-$MgB_6$, ort-$CaB_6$, hex-$MgB_6$, hex-$CaB_6$, and hex-$SrB_6$, the B 2*p* orbitals dominate the bands around the Fermi level, while the metal atomic orbitals have almost no component at the Fermi level. However, for hex-$ScB_6$, hex-$TiB_6$, and hex-$YB_6$, both B 2*p* and M *d* orbitals (M = Sc, Ti, Y) dominate the bands around the Fermi level. For ort-$MgB_6$, there are several Dirac cones above the Fermi level, close to the high-symmetry point

S in the first BZ, which is similar to that of the ort-$AlB_6$ [35]. There are many bands crossing the Fermi level, which forms five pockets at the FS. As shown in Fig. 3(a), there are two pockets centered at X and Y points, respectively, one pocket along Γ-X line, and two pockets along X-S line. For ort-$CaB_6$, there are one, three, one, and one band(s) crossing the Fermi level along Γ-X, X-S, S-Y, and Y-Γ, respectively. As shown in Fig. 3(b), these bands form one dumbbell-shaped pocket centered on X, one elliptical pocket centered on Y, and one pocket along X-S. For hex-$MgB_6$, at the K point, there is one Dirac cone across the Fermi level whose states are mainly from the 2*p* orbitals of the B kagome lattice. This phenomenon has also been observed in one recent theoretical work [61]. Actually, kagome 2D lattice has been extensively studied owing to its unique physics related to Dirac bands [72]. There are one, two, and one band(s) crossing the Fermi level along Γ-M, M-K, and K-Γ, respectively, which form a rectangular pocket centered on M and an elliptical pocket centered on K at the FS [see Fig. 3(c)]. For hex-$CaB_6$, there are one, one, and four bands crossing the Fermi level along Γ-M, M-K, and K-Γ, respectively, which is similar to those of the hex-$SrB_6$ [see Fig. 2(m)]. The FSs of hex-$CaB_6$ and hex-$SrB_6$ are almost the same. As shown in Fig. 3(d) and (g), there are one rectangular pocket centered on M and two pockets along K-Γ at the FS of both hex-$CaB_6$ and hex-$SrB_6$. For hex-$ScB_6$, hex-$TiB_6$, and hex-$YB_6$, there are one, one, and two band(s) crossing the Fermi level along Γ-M, M-K, and K-Γ, respectively, which are mainly dominated by B 2*p* and Sc-3*d*/Ti-3*d*/Y-4*d* orbitals. As shown in Fig. 3, the FSs of these three materials all include one hexagonal star shaped pocket centered on Γ and an elliptical pocket centered on K. The FSs of hex-$ScB_6$ and hex-$YB_6$ are almost the same, whose hexagonal star shaped pockets are larger than that of hex-$TiB_6$. From our calculated electronic DOSs, the amplitudes of the $N(E_F)$ for hex-$CaB_6$ and hex-$SrB_6$ are relatively larger than those of the other $MB_6$, which promotes the superconductivity of these three 2D materials.

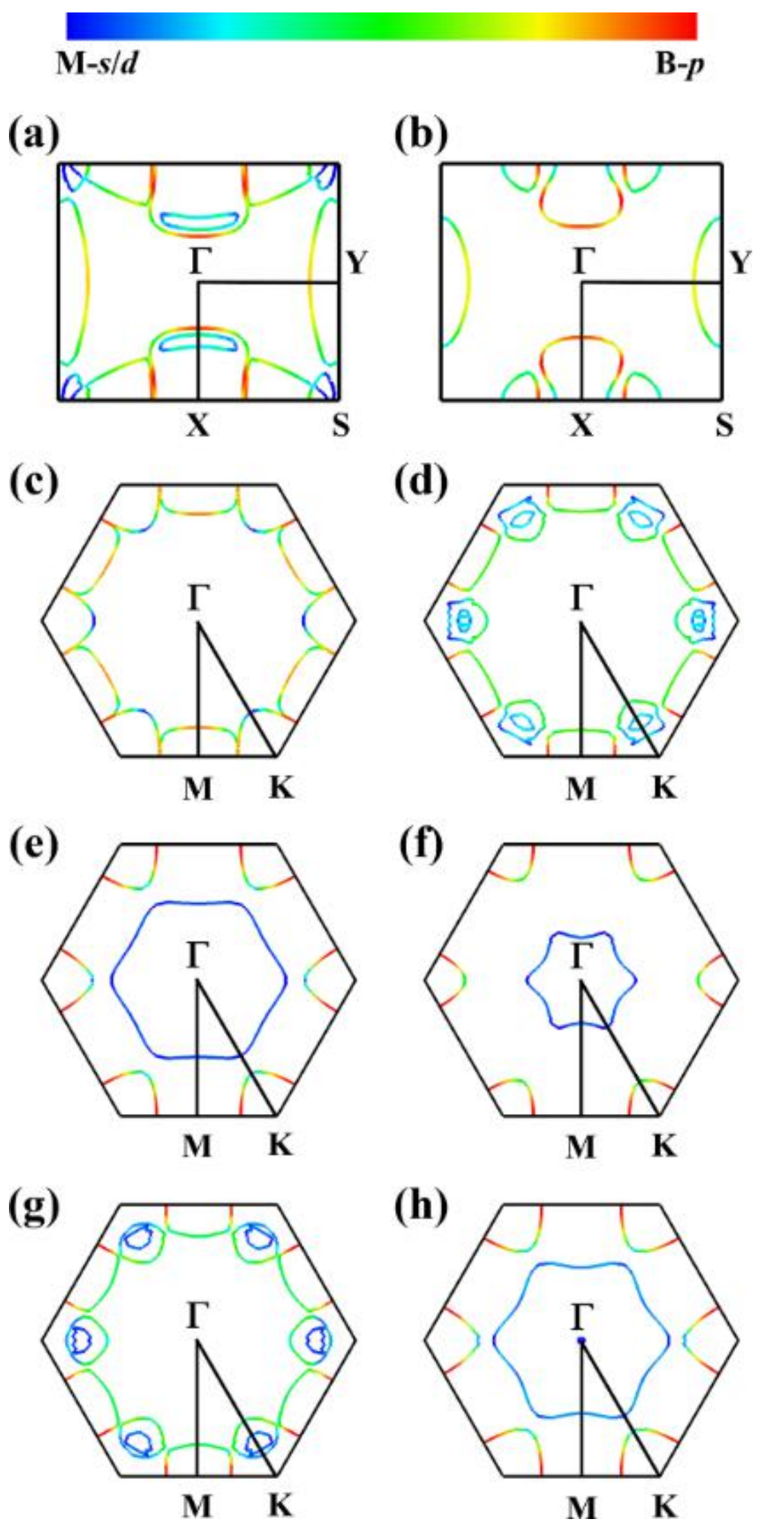

FIG. 3. Fermi surfaces of (a) ort-$MgB_6$, (b) ort-$CaB_6$, (c) hex-$MgB_6$, (d) hex-$CaB_6$, (e) hex-$ScB_6$, (f) hex-$TiB_6$, (g) hex-$SrB_6$, and (h) hex-$YB_6$.

### D. EPC and superconductivity

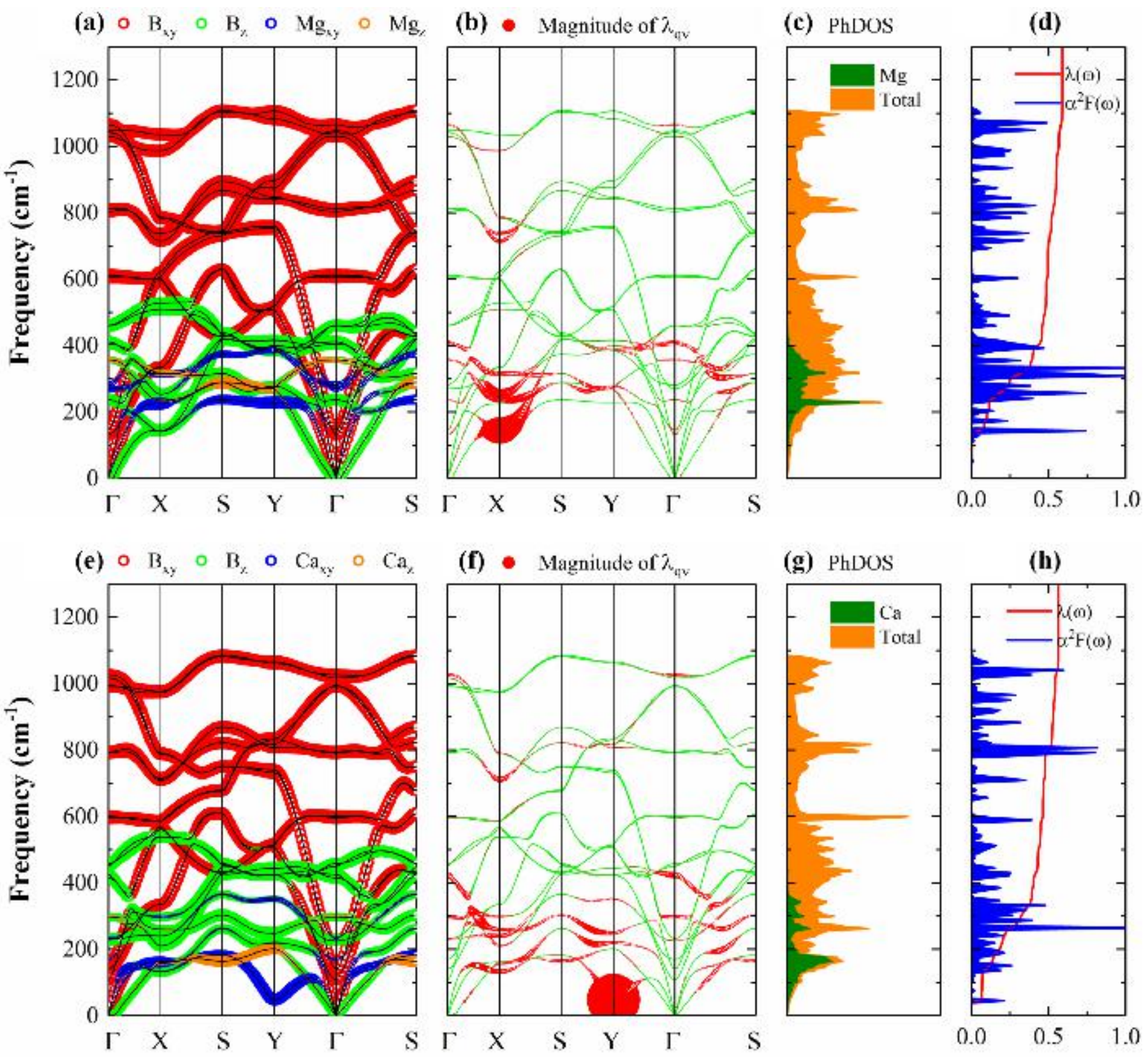

FIG. 4. Phonon dispersions, PhDOS, Eliashberg spectral function $\alpha^2F(\omega)$, and cumulative frequency dependence of EPC $\lambda(\omega)$ of (a-d) ort-$MgB_6$ and (e-h) ort-$CaB_6$. In the first-left panels, the phonon dispersions are weighted by the motion modes of B, M (M = Mg, Ca) atoms, while in the second-left panels, the phonon dispersions are weighted by the magnitude of EPC $\lambda_{qv}$. The area of the circles is proportional to the magnitude of EPC $\lambda_{qv}$. The red, green, blue, and orange hollow circles in (a) and (e) represent B horizontal, B vertical, M (M = Mg, Ca) horizontal, and M (M = Mg, Ca) vertical modes, respectively.

In order to investigate the EPC and superconducting transition temperature $T_c$ of these 2D $MB_6$, we calculate their phonon dispersions, PhDOS, Eliashberg spectral function $\alpha^2F(\omega)$, and $\lambda(\omega)$ and present them in Figs. 4, 5, and 7. As shown in Fig. 4 and Fig. 5, there are no imaginary phonon modes in the phonon dispersions for ort-$MgB_6$, ort-$CaB_6$, hex-$MgB_6$, hex-$CaB_6$, hex-$ScB_6$, and hex-$TiB_6$. This clearly indicates that these materials are dynamically stable. The EPC integrated over all phonon branches and distributed in the BZ are shown in Fig. 6 for these six materials.

Our calculated phonon spectra for the ort-$MgB_6$ are consistent with those presented in Liao's work [32]. From our decomposed phonon dispersions, we find that the Mg vibrations and the out-of-plane $B_z$ vibrations dominate the low frequencies below 500 $cm^{-1}$, while the in-plane $B_{xy}$ vibrations dominate the high frequencies above 500 $cm^{-1}$. The low-frequency phonons (150-350 $cm^{-1}$) around X point, mainly associated with the $B_z$ vibrations, contribute 68% of the total EPC ($\lambda$=0.56). As shown in Fig. 6(a), the region around the X point possesses the largest EPC, which is consistent with the EPC $\lambda_{qv}$ in Fig. 4(b). The phonons (300-400 $cm^{-1}$) along the S-Y-Γ-S line, also mainly associated with the $B_z$ vibrations, contribute 21% of the total EPC ($\lambda$=0.56). Finally, we obtain the $T_c$ = 8.8 K for the ort-$MgB_6$ with $\omega_{\log}$ = 467.6 K. Our calculated $\lambda$=0.56, $\omega_{\log}$ = 467.6 K, and $T_c$ = 8.8 K of the ort-$MgB_6$ are comparable to those of Liao' work ($\lambda$=0.49, $\omega_{\log}$ = 554.2 K, and $T_c$ = 6.0 K) [32]. For ort-$CaB_6$, the Ca vibrations mainly dominate the low frequencies of 0-200 $cm^{-1}$, while the B vibrations spread in whole area of BZ. Surprisingly, there is a softened phonon mode around the Y point, associated with the $Ca_{xy}$ vibration, contribute 14% of the total EPC ($\lambda$=0.57). The phonons (100-400 $cm^{-1}$) in the full BZ, mainly associated with the $B_z$ vibration, contribute 56% of the total EPC ($\lambda$=0.57). As shown in Fig. 6(b), the largest EPC appears at Y point and the EPC spreads in the full BZ. The calculated $T_c$ is 7.1 K, which is comparable to that of the ort-$MgB_6$. The EPC values of 0.56 and 0.57 indicate that both ort-$MgB_6$ and ort-$CaB_6$ belong to weak conventional superconductors.

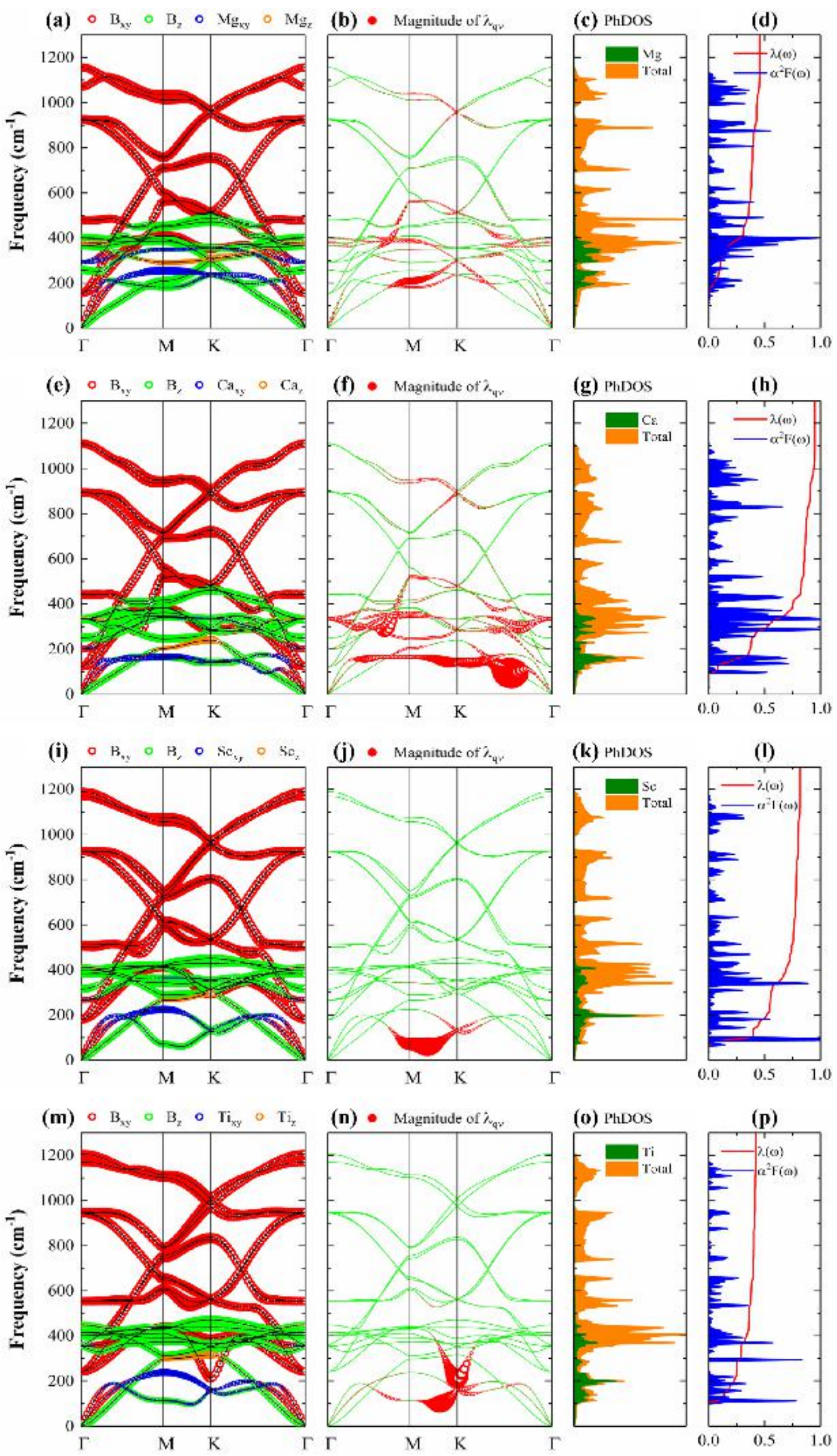

FIG. 5. Phonon dispersions, PhDOS, Eliashberg spectral function $\alpha^2F(\omega)$, and cumulative frequency dependence of EPC $\lambda(\omega)$ of (a-d) hex-$MgB_6$, (e-h) hex-$CaB_6$, (i-l) hex-$ScB_6$, and (m-p) hex-$TiB_6$. The details and colors are the same as in Fig. 4.

For hex-$MgB_6$, hex-$CaB_6$, hex-$ScB_6$, and hex-$TiB_6$, as shown in Fig. 5, the $B_z$ vibrations are mainly distributed in the frequency range of 0-500 $cm^{-1}$, while the $B_{xy}$ vibrations is distributed in whole area of the BZ. For hex-$MgB_6$, the Mg vibrations are mainly distributed in the frequency range of 200-400 $cm^{-1}$. About 75% of the total

EPC ($\lambda$=0.46) arises from the phonon frequency range of ~180-400 $cm^{-1}$, which is mainly from the $B_z$ vibrations. As shown in Fig. 6(c), the region with the largest EPC is around the M point, and there is also strong EPC in the region around the K point. However, in the region around the Γ point, the EPC is very small. These results are consistent with the EPC $\lambda_{qv}$ in Fig. 5(b). For hex-$CaB_6$, the Ca vibrations are mainly distributed in the low frequency below 210 $cm^{-1}$. There are two softened acoustic branches with the phonon frequency below 150 $cm^{-1}$, which are mainly from the $Ca_{xy}$ vibrations. The softened phonon modes result in a strong EPC, which account for 39% the total EPC ($\lambda$=0.94). The phonon modes in the frequency range of 200-500 $cm^{-1}$, mainly associated with the $B_z$ vibrations, contribute 46% to the total EPC ($\lambda$=0.94). As shown in Fig. 6(d), the region with the largest EPC is along the Γ-K line, and there is strong EPC in the whole BZ, which is consistent with the EPC $\lambda_{qv}$ in Fig. 5(f). For hex-$ScB_6$, the Sc vibrations are mainly distributed in the frequency range of 150-300 $cm^{-1}$. There are two softened acoustic branches, one is softening along the M-K line, the other around the K point. About 70% of the total EPC ($\lambda$=0.81) arises from these two softened phonon modes, which are mainly from the $B_z$ vibrations. From the distribution map of EPC [see Fig. 6(e)], we find that the region with the largest EPC is along the M-K line and there is also strong EPC around the K point, which is consistent with the EPC $\lambda_{qv}$ in Fig. 5(j). For hex-$TiB_6$, the Ti vibrations are mainly distributed in the frequency range of 100-300 $cm^{-1}$. The softened phonon modes (100-300 $cm^{-1}$), mainly associated with the $B_{xy}$ and $B_z$ vibrations, contribute about 68% of the total EPC ($\lambda$=0.44). The EPC is mainly in the region around the K point, which can be verified by the distribution map of EPC shown in Fig. 6(f). The EPC values of 0.46, 0.94, 0.81, and 0.44 indicate that the hex-$MgB_6$ and hex-$TiB_6$ belong to weak conventional superconductors and the hex-$CaB_6$ and hex-$ScB_6$ belong to moderate conventional superconductors. Then, the calculated $T_c$ are 5.0, 21.3, 7.7, and 2.2 K for hex-$MgB_6$, hex-$CaB_6$, hex-$ScB_6$, and hex-$TiB_6$, respectively.

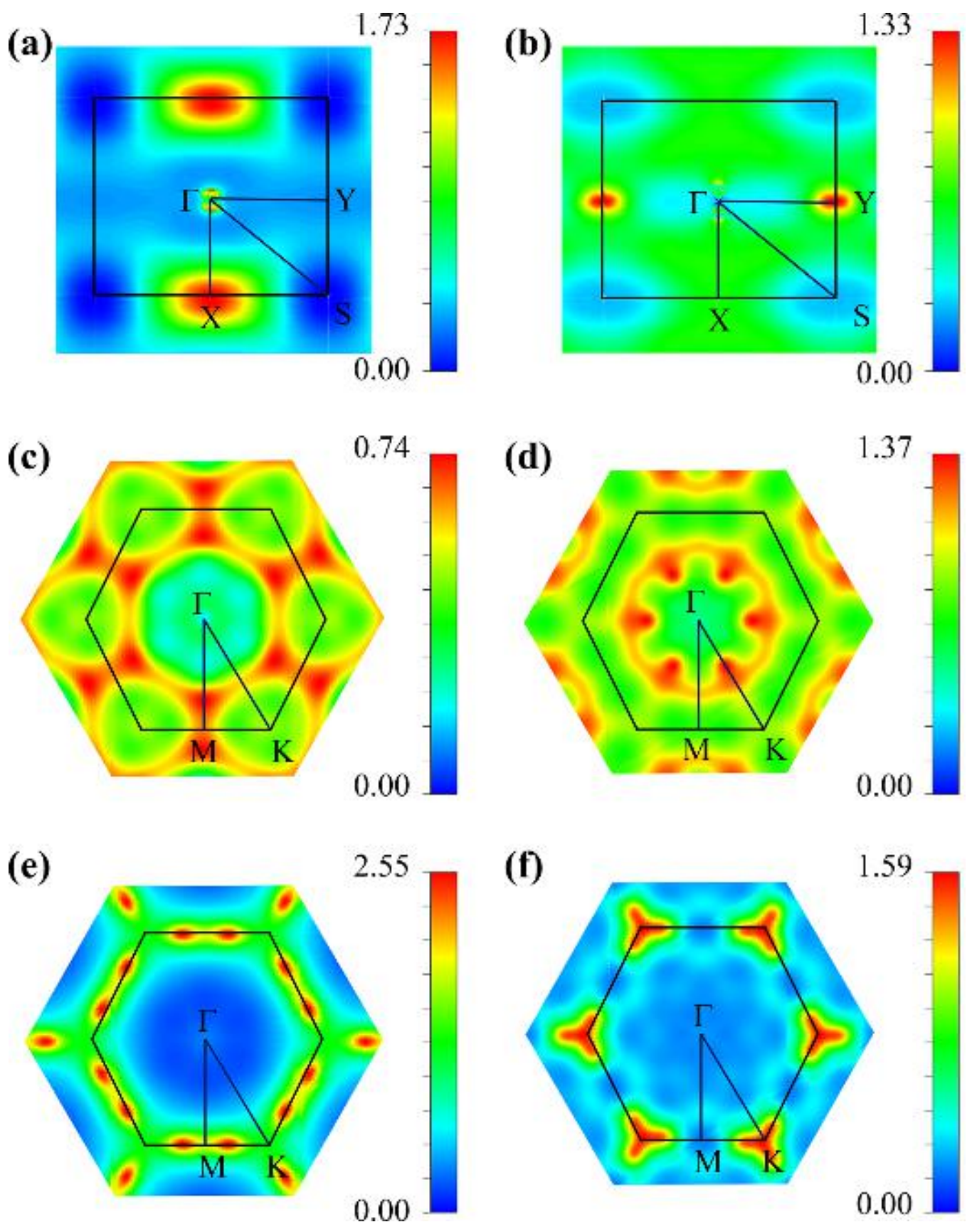

FIG. 6. The integrated EPC distributions in the BZs for (a) ort-$MgB_6$, (b) ort-$CaB_6$, (c) hex-$MgB_6$, (d) hex-$CaB_6$, (e) hex-$ScB_6$, and (f) hex-$TiB_6$.

For hex-$SrB_6$ and hex-$YB_6$, as shown in Fig. 7, there exist imaginary frequencies in their BZ. However, such small negative vibrations can be understood by the Kohn anomaly [73]. The Kohn anomaly of the phonon mode always yields significant coupling between electrons and acoustic phonons and may induce large EPC [24, 31]. In our study of hex-$SrB_6$ and hex-$YB_6$, the Kohn anomalies along Γ-K and M-K do exhibit large EPC. For these two materials, the $B_z$ vibrations are mainly distributed in the frequency range below 500 cm$^{-1}$, while the $B_{xy}$ vibrations is distributed in whole area of the BZ. For hex-$SrB_6$, the Kohn anomaly along the K to Γ point is caused by $Sr_{xy}$ vibrations. The phonon modes (below 100 cm$^{-1}$), dominated by both Sr and B modes, contribute about 65% of the total EPC ($\lambda$=1.71). The rest of the EPC (35%) mainly comes from the phonon modes (200-500 cm$^{-1}$) from $B_{xy}$ and $B_z$ vibrations. For hex-$YB_6$, the Kohn anomaly along the M point to K point is mainly from the $B_z$ vibrations. About 83% of the total EPC ($\lambda$=1.82) arises from the phonon frequency range below 150 $cm^{-1}$, which is mainly from the $B_z$ and $Y_{xy}$ vibrations. The rest of the EPC (17%) mainly comes from the phonon modes (300-500 $cm^{-1}$) from $B_{xy}$ and $B_z$ vibrations. The EPC values of 1.71 and 1.82 indicate that both hex-$SrB_6$ and hex-$YB_6$ belong to strong conventional superconductors. The calculated $T_c$ are 15.9 and 11.5 K for hex-$SrB_6$ and hex-$YB_6$, respectively.

In Table III, we list the superconducting parameters of $\mu^*$, $N(E_F)$, $\omega_{log}$, EPC constant $\lambda$ and $T_c$ for our calculated eight 2D $MB_6$ and other intrinsic borophenes [24, 26] and 2D metal borides superconductors [32-35, 45]. Among the 2D $MB_6$ we studied, hex-$CaB_6$ and hex-$SrB_6$ have the largest $N(E_F)$ and possess the largest values of $T_c$. This indicates that the occupation of electrons at the Fermi level has a positive effect on superconductivity. The EPC constant $\lambda$ is in the range of 0.44 ~ 1.82 and the $T_c$ is in the range of 2.2 ~ 21.3 K. While the hex-$TiB_6$ has the smallest $T_c$ (2.2 K) with the smallest EPC constant ($\lambda$=0.44), the hex-$CaB_6$ possesses the largest $T_c$ of 21.3 K. Compared with hex-$CaB_6$, hex-$SrB_6$ and hex-$YB_6$ have larger EPC constant ($\lambda$=1.71 and 1.82, respectively), but hold smaller $T_c$. This is mainly caused by their small $\omega_{log}$. To have a more comprehensive view of our predicted 2D $MB_6$, we compare their EPC constants $\lambda$ and $T_c$ with the reported borophenes [24, 26] and 2D metal borides [32-35, 45]. Among all the materials listed in Table III, our studied hex-$CaB_6$ possesses the largest $T_c$ (21.3 K). In addition to hex-$CaB_6$, our obtained hex-$SrB_6$ also have large $T_c$ of 15.9 K, which is larger than that of $B_\diamond$ [24], $\beta_{12}$ [26], ort-$AlB_6$ [35], ort-/hex-$GaB_6$ [45], ort-/hex-$InB_6$ [45], $Mo_2B_2$ [33], and $W_2B_2$ [34], while smaller than that of $B_\triangle$/$B_\square$ [24]. The $T_c$ of our studied materials, except for hex-$TiB_6$, is greater than that of ort-$AlB_6$ [35]. Compared with bulk $YB_6$ [44, 74], 2D hex-$YB_6$ has larger $N(E_F)$ and EPC constant, which leads to a larger $T_c$. The different arrangement of atoms in the crystal cell results in the difference of superconductivity between 2D and bulk $YB_6$.

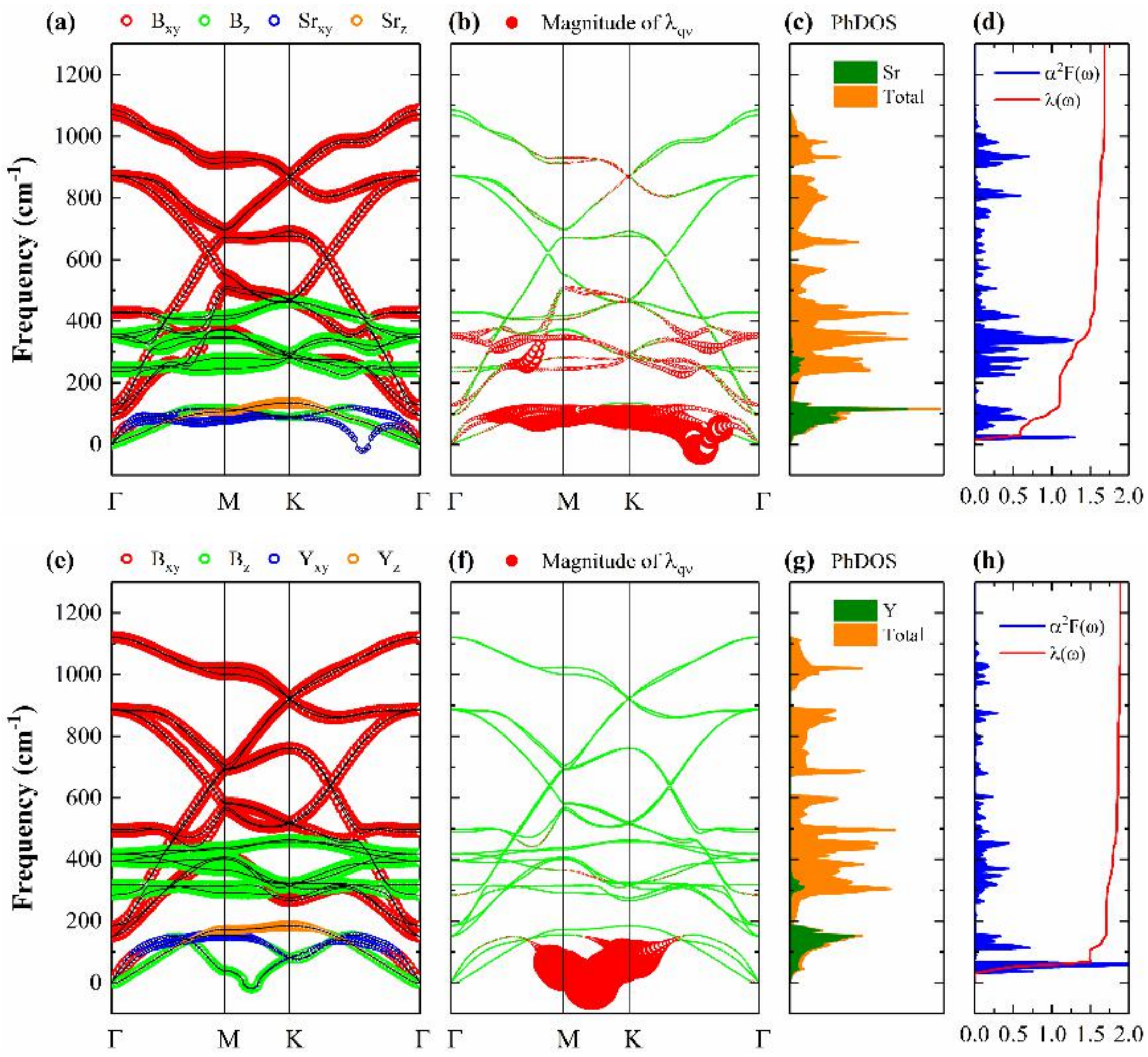

FIG. 7. Phonon dispersions, PhDOS, Eliashberg spectral function $\alpha^2F(\omega)$, and cumulative frequency dependence of EPC $\lambda(\omega)$ of (a-d) hex-$SrB_6$ and (e-h) hex-$YB_6$. The details and colors are the same as in Fig. 4.

TABLE III. Superconducting parameters of $\mu^*$, $N(E_F)$ (States/spin/Ry/Unit Cell), $\omega_{log}$ (K), EPC constant $\lambda$ and $T_c$ (K) for our calculated 2D $MB_6$ and other intrinsic borophenes and 2D metal borides superconductors. The theoretical and experimental data of bulk *Pm*-3*m*-$YB_6$ is also presented.

| Materials | $\mu^*$ | $N(E_F)$ | $\omega_{log}$ | λ | $T_c$ | Refs. |
|---|---|---|---|---|---|---|
| Ort-$MgB_6$ | 0.1 | 8.9 | 467.6 | 0.56 | 8.8 | This work |
| Ort-$CaB_6$ | 0.1 | 9.8 | 374.7 | 0.57 | 7.1 | This work |
| Hex-$MgB_6$ | 0.1 | 7.3 | 561.5 | 0.46 | 5.0 | This work |
| Hex-$CaB_6$ | 0.1 | 11.7 | 416.4 | 0.94 | 21.3 | This work |
| Hex-$ScB_6$ | 0.1 | 7.3 | 269.2 | 0.81 | 7.7 | This work |
| Hex-$TiB_6$ | 0.1 | 6.5 | 323.0 | 0.44 | 2.2 | This work |
| Hex-$SrB_6$ | 0.1 | 13.9 | 130.0 | 1.71 | 15.9 | This work |
| Hex-$YB_6$ | 0.1 | 8.4 | 101.9 | 1.82 | 11.5 | This work |
| $B_{\triangle}/B_{\square}/B_{\diamond}$ | 0.1 | | | 1.1/0.8/0.6 | 21/16/12 | [24] |
| $\beta_{12}$ | 0.1 | | | 0.69 | 14 | [26] |
| Ort-$MgB_6$ | 0.1 | | 554.2 | 0.49 | 6.0 | [32] |
| Ort-$AlB_6$ | 0.05 | 6.41 | 345.2 | 0.36 | 4.7 | [35] |
| Ort-$GaB_6$ | 0.1 | 6.40 | 490.3 | 0.39 | 1.67 | [45] |
| Hex-$GaB_6$ | 0.1 | 7.79 | 436.1 | 0.68 | 14.02 | [45] |
| Ort-$InB_6$ | 0.1 | 8.69 | 471.2 | 0.55 | 7.77 | [45] |
| Hex-$InB_6$ | 0.1 | 9.43 | 154.1 | 0.67 | 4.83 | [45] |
| Tetr-$Mo_2B_2$ | 0.1 | 16.02 | 344.84 | 0.49 | 3.9 | [33] |
| Tri-$Mo_2B_2$ | 0.1 | 16.81 | 295.0 | 0.3 | 0.2 | [33] |
| Tetr-$W_2B_2$ | 0.1 | 12.46 | 232.4 | 0.69 | 7.8 | [34] |
| Hex-$W_2B_2$ | 0.1 | 13.60 | 232.2 | 0.43 | 1.5 | [34] |
| *Pm*-3*m*-$YB_6$ | 0.1 | 1.36 | 72.31 | 0.87 | 7.47 | [74] |
| [a]*Pm*-3*m*-$YB_6$ | | | | | 7.2 | [44] |

[a] This is experimental data.

## E. Strain engineering

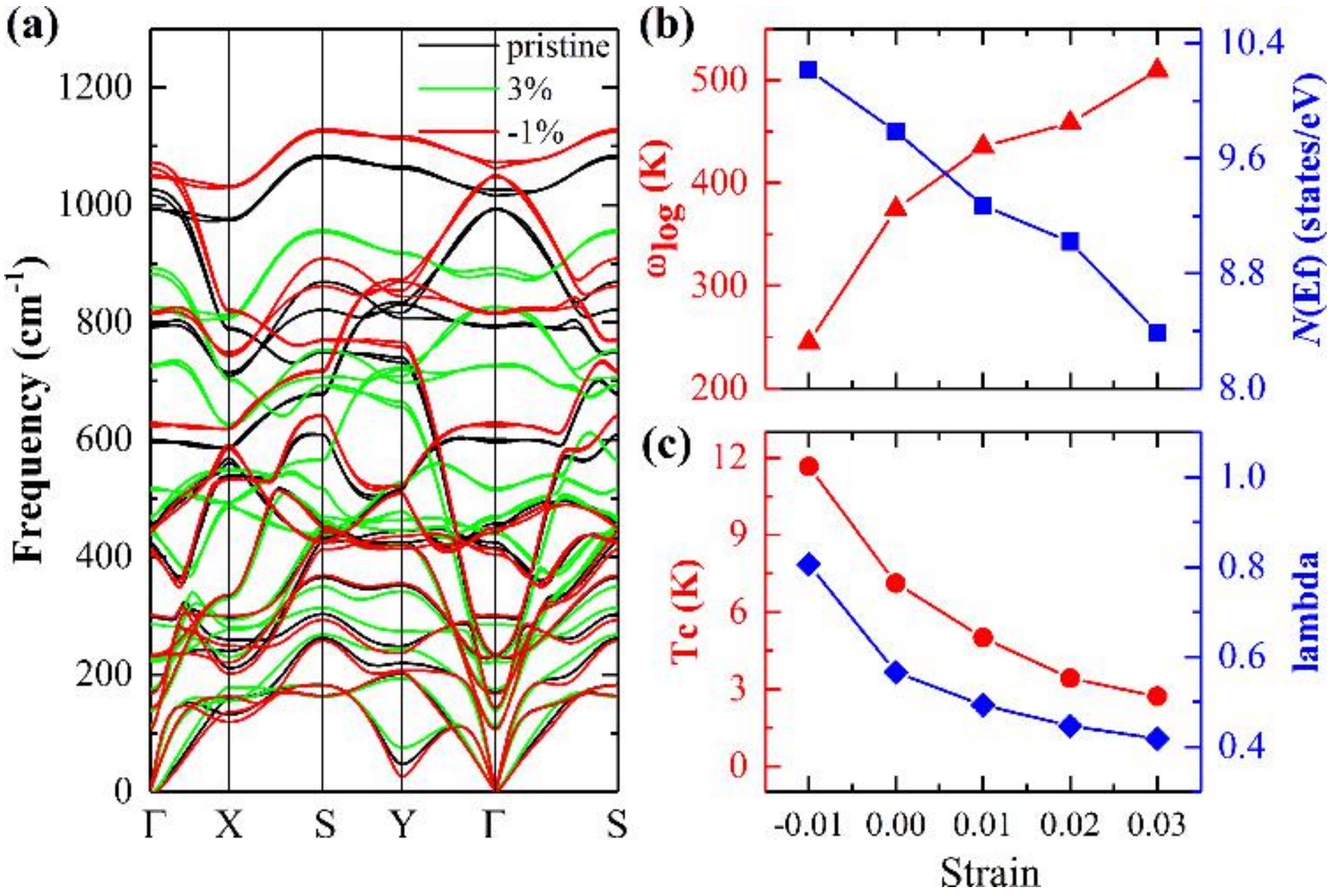

FIG. 8. (a) Phonon dispersions, (b) logarithmically averaged phonon frequency $\omega_{\log}$ (red) and $N(E_F)$ (blue), (c) $T_c$ (red) and EPC constant $\lambda$ (blue) under different strains for ort-$CaB_6$.

When these 2D $MB_6$ are grown on substrates, strains will be introduced, therefore, the superconducting properties should be affected. Therefore, we apply strain by changing the lattice parameters to investigate the effect of strains on the EPC and superconducting properties of these 2D materials. Here, we choose the ort-$CaB_6$ and the hex-$CaB_6$ as the representatives to study. First, we discuss the dynamical stability of the ort-$CaB_6$ and the hex-$CaB_6$ under different strains. The equi-biaxial tensile and compressive strains have been applied on these two configurations. The positive value of $\varepsilon$ represents tensile strain while the negative value represents compressive strain. For ort-$CaB_6$, we find that it is dynamically stable under the strain region of $-1\% \leq \varepsilon \leq 3\%$, indicating that it can withstand up to 1% compressive strain and 3% tensile strain. The phonon dispersions under $\varepsilon$ = -1%, 0, and 3% are displayed in Fig. 8(a). We find that the compression strain makes the phonon frequency at low frequency (< 500 $cm^{-1}$) lower down, but those at high frequency (> 500 $cm^{-1}$) higher up. When tensile strain is applied, the trends of the phonon frequency change are opposite. As shown in Fig. 8(b) and (c), when the strain $\varepsilon$ increases from -1% to 3%, the logarithmically averaged phonon frequency $\omega_{\log}$ increases linearly, whereas the $N(E_F)$, the EPC constant $\lambda$, and the $T_c$ decreases linearly. Under compressive strain of -1%, the largest $T_c$ of 12 K is obtained.

For hex-$CaB_6$, it is dynamically stable under the strain region of -3% $\leq \varepsilon \leq$ 2%, indicating that it can bear more compression strain than ort-$CaB_6$. The phonon dispersions under $\varepsilon$ = -3%, -2%, 0, and 2% are displayed in Fig. 9(a). Like ort-$CaB_6$, the compression strain makes the phonon frequency at low frequency (< 500 $cm^{-1}$) lower down, but those at high frequency (> 500 $cm^{-1}$) higher up. As shown in Fig. 9(b) and (c), when the strain $\varepsilon$ increases from -3% to 2%, the logarithmically averaged phonon frequency $\omega_{log}$ increases linearly, whereas the $N(E_F)$, the EPC constant $\lambda$, and the $T_c$ are almost linearly reduced. Under the maximum compressive strain of $\varepsilon$ = -3%, the $T_c$ can be modulated up to 28 K. Therefore, the compressive strain can strengthen the EPC and increase the $T_c$ of both ort-$CaB_6$ and hex-$CaB_6$.

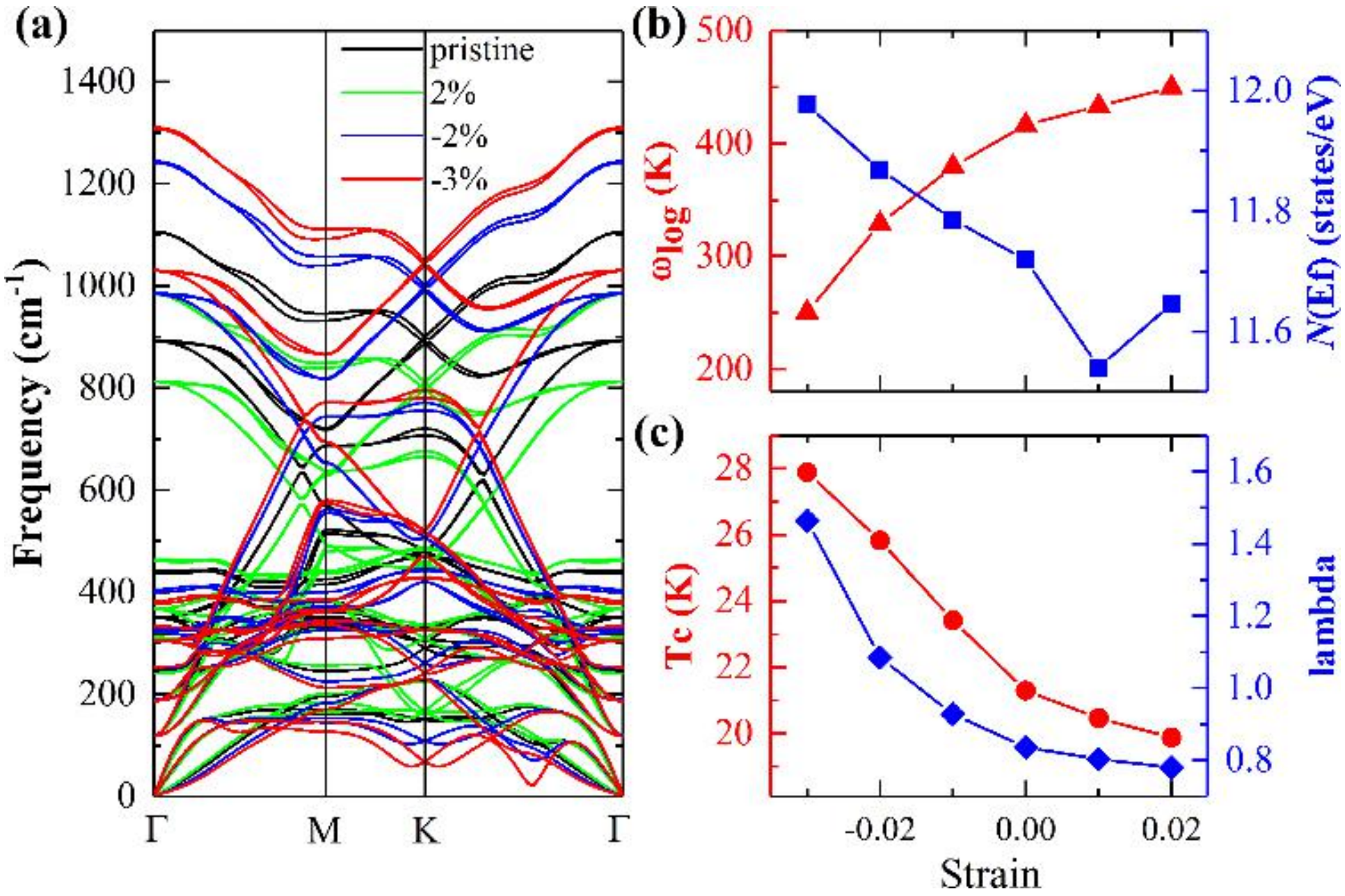

FIG. 9. (a) Phonon dispersions, (b) logarithmically averaged phonon frequency $\omega_{log}$ (red) and $N(E_F)$ (blue), (c) $T_c$ (red) and EPC constant $\lambda$ (blue) under different strains for hex-$CaB_6$.

## IV. CONCLUSION

In our present work, by combining the crystal structure search as well as systematic first-principles calculations, we predict a series of 2D metal borides, including orthorhombic $MB_6$ (M=Mg, Ca) and hexagonal $MB_6$ (M=Mg, Ca, Sc, Ti, Sr, Y), and investigate their geometrical structures, bonding properties, electronic structures, mechanical properties, phonon dispersions, dynamic stability, EPC, superconducting properties and so on. These ort-$MB_6$ and hex-$MB_6$ structures are subjected to the orthogonal *Pmmm* and the hexagonal *P6/mmm* space group, respectively, which consist one layer of metal atoms and two layers of B atoms. Some of these materials contain covalent and ionic bonds, while others also have metal bonds in addition to the above two bonds. The relatively large cohesive energies indicate that these $MB_6$ monolayers may be synthesized under appropriate experimental conditions. Our AIMD simulations show that these 2D $MB_6$ well maintain their original configurations up to 700/1000 K, indicating the excellent thermodynamic stability of these $MB_6$ monolayers. For the ort-/hex-$MgB_6$, ort-/hex-$CaB_6$, hex-$ScB_6$, and hex-$TiB_6$, no imaginary modes in the entire BZ proves that they are dynamically stable. While for the hex-$SrB_6$ and hex-$YB_6$, there exist small imaginary frequencies standing for the Kohn anomaly. Our calculated elastic constants clearly illustrate that that these $MB_6$ monolayers are all mechanically stable. The strong B-B bonding enable large values of $C_{11}$. Thus, once our predicted 2D $MB_6$ being synthesized in experiments, good stability is expected.

According to our EPC calculations based upon the BCS theory, these eight 2D $MB_6$ materials are all intrinsic phonon-mediated superconductors with the $T_c$ in the range of 2.2 to 21.3 K. Among them, there are six weak conventional superconductors with small EPC constants $\lambda$ (0.44-0.94), including ort-$MgB_6$, ort-$CaB_6$, hex-$MgB_6$, hex-$TiB_6$, hex-$CaB_6$, and hex-$ScB_6$, while there are two strong conventional superconductors with large EPC constants $\lambda$ (1.71-1.82), including hex-$SrB_6$ and hex-$YB_6$. The EPC is mainly governed by the B vibrations while in little cases some couplings from the metal atoms. The soft modes as well as the Kohn anomaly for acoustic phonon branches play critical role for the large EPC. We also study the effects of tensile/compressive strains on the superconducting properties of the ort-$CaB_6$ and hex-$CaB_6$. Results show that compressive strains obviously enhance the EPC and increase the $T_c$ for both ort-$CaB_6$ and hex-$CaB_6$, which is due to the lower acoustic phonon branch is softened. A large $T_c$ of 28 K for hex-$CaB_6$ is obtained.

Our findings in the present work enrich the database of 2D metal borides and may stimulate experimental synthesis of these materials and also 2D metal carbides and nitrides. The superconducting properties of these 2D $MB_6$ supply experiments another choice other than the 2D boron to realize superconducting monolayers in boron based 2D forms. After all, the thickness of these 2D $MB_6$ is prior to borophene and their $T_c$ are relatively high.

## NOTES

The authors declare no competing financial interest.

## ACKNOWLEDGEMENTS

We acknowledge financial support from Guangdong Basic and Applied Basic Research Foundation under Grant No. 2019A1515110920 and the PhD Start-up Fund of the Natural Science Foundation of Guangdong Province of China (Grant No. 2018A0303100013). The work was completed with the computational resources from the Supercomputer Centre of the China Spallation Neutron Source.

## SUPPLEMENTAL MATERIALS

See Supplemental Material at [URL will be inserted by publisher] for (I) Electron localization function (ELF) of the $MB_6$ monolayers, and (II) snapshots and variation of the free energy for the 2D $MB_6$ in the AIMD simulations from 300 to 1000 K.

## REFERENCES

[1] E. Dagotto, Rev. Mod. Phys. **66**, 763 (1994).

[2] P. Dai, Rev. Mod. Phys. **87**, 855 (2015).

[3] F. Giustino, Rev. Mod. Phys. **89**, 015003 (2017).

[4] J. Nagamatsu, N. Nakagawa, T. Muranaka, Y. Zenitani, and J. Akimitsu, Nature **410**, 63 (2001).

[5] A. Y. Liu, I. I. Mazin, and J. Kortus, Phys. Rev. Lett. **87**, 087005 (2001).

[6] A. Floris, G. Profeta, N. N. Lathiotakis, M. Lüders, M. A. L. Marques, C. Franchini, E. K. U.

Gross, A. Continenza, and S. Massidda, Phys. Rev. Lett. **94**, 037004 (2005).
[7] E. R. Margine and F. Giustino, Phys. Rev. B **87**, 024505 (2013).
[8] A. Bhaumik, R. Sachan, and J. Narayan, ACS Nano **11**, 5351 (2017).
[9] X. Liang, A. Bergara, L. Wang, B. Wen, Z. Zhao, X.-F. Zhou, J. He, G. Gao, and Y. Tian, Phys. Rev. B **99**, 100505(R) (2019).
[10] Y. Li, J. Hao, H. Liu, Y. Li, and Y. Ma, J. Chem. Phys. **140**, 174712 (2014).
[11] H. Liu, I. I. Naumov, R. Hoffmann, N. W. Ashcroft, and R. J. Hemley, P. Natl. Acad. Sci. **114**, 6990 (2017).
[12] J. Bardeen, L. N. Cooper, and J. R. Schrieffer, Phys. Rev. **108**, 1175 (1957).
[13] A. P. Drozdov, M. I. Eremets, I. A. Troyan, V. Ksenofontov, and S. I. Shylin, Nature **525**, 73 (2015).
[14] A. P. Drozdov *et al.*, Nature **569**, 528 (2019).
[15] M. Somayazulu, M. Ahart, A. K. Mishra, Z. M. Geballe, M. Baldini, Y. Meng, V. V. Struzhkin, and R. J. Hemley, Phys. Rev. Lett. **122**, 027001 (2019).
[16] T. Uchihashi, Supercond. Sci. Tech. **30**, 013002 (2016).
[17] C. Brun, T. Cren, and D. Roditchev, Supercond. Sci. Tech. **30**, 013003 (2016).
[18] C. Xu, L. Wang, Z. Liu, L. Chen, J. Guo, N. Kang, X.-L. Ma, H.-M. Cheng, and W. Ren, Nat. Mater. **14**, 1135 (2015).
[19] B. M. Ludbrook *et al.*, P. Natl. Acad. Sci. **112**, 11795 (2015).
[20] X. Xi, Z. Wang, W. Zhao, J.-H. Park, K. T. Law, H. Berger, L. Forró, J. Shan, and K. F. Mak, Nat. Phys. **12**, 139 (2016).
[21] L. Zhu, Q.-Y. Li, Y.-Y. Lv, S. Li, X.-Y. Zhu, Z.-Y. Jia, Y. B. Chen, J. Wen, and S.-C. Li, Nano Lett. **18**, 6585 (2018).
[22] M. Liao *et al.*, Nat. Phys. **14**, 344 (2018).
[23] Y. Cao, V. Fatemi, S. Fang, K. Watanabe, T. Taniguchi, E. Kaxiras, and P. Jarillo-Herrero, Nature **556**, 43 (2018).
[24] E. S. Penev, A. Kutana, and B. I. Yakobson, Nano Lett. **16**, 2522 (2016).
[25] M. Gao, Q.-Z. Li, X.-W. Yan, and J. Wang, Phys. Rev. B **95**, 024505 (2017).
[26] C. Cheng, J.-T. Sun, H. Liu, H.-X. Fu, J. Zhang, X.-R. Chen, and S. Meng, 2D Mater. **4**, 025032 (2017).
[27] M. Gao, X.-W. Yan, J. Wang, Z.-Y. Lu, and T. Xiang, Phys. Rev. B **100**, 024503 (2019).
[28] H.-Y. Lu *et al.*, Phys. Rev. B **101**, 214514 (2020).
[29] E. R. Margine and F. Giustino, Phys. Rev. B **90**, 014518 (2014).
[30] C. Si, Z. Liu, W. Duan, and F. Liu, Phys. Rev. Lett. **111**, 196802 (2013).
[31] B.-T. Wang, P.-F. Liu, T. Bo, W. Yin, O. Eriksson, J. Zhao, and F. Wang, Phys. Chem. Chem. Phys. **20**, 12362 (2018).
[32] J.-H. Liao, Y.-C. Zhao, Y.-J. Zhao, H. Xu, and X.-B. Yang, Phys. Chem. Chem. Phys. **19**, 29237 (2017).
[33] L. Yan, T. Bo, P.-F. Liu, B.-T. Wang, Y.-G. Xiao, and M.-H. Tang, J. Mater. Chem. C **7**, 2589 (2019).
[34] L. Yan, T. Bo, W. Zhang, P.-F. Liu, Z. Lu, Y.-G. Xiao, M.-H. Tang, and B.-T. Wang, Phys. Chem. Chem. Phys. **21**, 15327 (2019).
[35] B. Song, Y. Zhou, H.-M. Yang, J.-H. Liao, L.-M. Yang, X.-B. Yang, and E. Ganz, J. Am. Chem. Soc. **141**, 3630 (2019).

[36] B.-T. Wang, P.-F. Liu, J.-J. Zheng, W. Yin, and F. Wang, Phys. Rev. B **98**, 014514 (2018).
[37] F. Weber *et al.*, Phys. Rev. Lett. **107**, 107403 (2011).
[38] J. Bekaert, M. Petrov, A. Aperis, P. M. Oppeneer, and M. V. Milošević, Phys. Rev. Lett. **123**, 097001 (2019).
[39] J.-J. Zhang, B. Gao, and S. Dong, Phys. Rev. B **93**, 155430 (2016).
[40] J.-J. Zhang, Y. Zhang, and S. Dong, Phys. Rev. Mater. **2**, 126004 (2018).
[41] C. Cheng, J.-T. Sun, M. Liu, X.-R. Chen, and S. Meng, Phys. Rev. Mater. **1**, 074804 (2017).
[42] J.-J. Zhang and S. Dong, J. Chem. Phys. **146**, 034705 (2017).
[43] L. Zhang, Y. Wang, J. Lv, and Y. Ma, Nat. Rev. Mater. **2**, 17005 (2017).
[44] R. Lortz *et al.*, Phys. Rev. B **73**, 024512 (2006).
[45] L. Yan, T. Bo, P.-F. Liu, L. Zhou, J. Zhang, M.-H. Tang, Y.-G. Xiao, and B.-T. Wang, J. Mater. Chem. C **8**, 1704 (2020).
[46] Y. Wang, J. Lv, L. Zhu, and Y. Ma, Phys. Rev. B **82**, 094116 (2010).
[47] Y. Wang, J. Lv, L. Zhu, and Y. Ma, Comput. Phys. Commun. **183**, 2063 (2012).
[48] Y. Wang, M. Miao, J. Lv, L. Zhu, K. Yin, H. Liu, and Y. Ma, J. Chem. Phys. **137**, 224108 (2012).
[49] G. Kresse and J. Furthmüller, Phys. Rev. B **54**, 11169 (1996).
[50] J. P. Perdew, K. Burke, and M. Ernzerhof, Phys. Rev. Lett. **77**, 3865 (1996).
[51] P. E. Blöchl, Phys. Rev. B **50**, 17953 (1994).
[52] J. Klimeš, D. R. Bowler, and A. Michaelides, Phys. Rev. B **83**, 195131 (2011).
[53] M. Fuchs and M. Scheffler, Comput. Phys. Commun. **119**, 67 (1999).
[54] P. Giannozzi *et al.*, J. Phys-Condens. Mat. **21**, 395502 (2009).
[55] S. Baroni, S. de Gironcoli, A. Dal Corso, and P. Giannozzi, Rev. Mod. Phys. **73**, 515 (2001).
[56] F. Giustino, Rev. Mod. Phys. **89**, 015003 (2017).
[57] B. Luo *et al.*, P. Natl. Acad. Sci. USA **116**, 17213 (2019).
[58] X.-H. Tu, P.-F. Liu, and B.-T. Wang, Phys. Rev. Mater. **3**, 054202 (2019).
[59] Y. Sun, J. Lv, Y. Xie, H. Liu, and Y. Ma, Phys. Rev. Lett. **123**, 097001 (2019).
[60] X. Wu, J. Dai, Y. Zhao, Z. Zhuo, J. Yang, and X. C. Zeng, ACS Nano **6**, 7443 (2012).
[61] S.-Y. Xie, X.-B. Li, W. Q. Tian, N.-K. Chen, Y. Wang, S. Zhang, and H.-B. Sun, Phys. Chem. Chem. Phys. **17**, 1093 (2015).
[62] A. Savin, R. Nesper, S. Wengert, and T. F. Fässler, Angew. Chem. Int. Edit. **36**, 1808 (1997).
[63] W. Tang, E. Sanville, and G. Henkelman, J. Phys-Condens. Mat. **21**, 084204 (2009).
[64] L. M. Yang, V. Bacic, I. A. Popov, A. I. Boldyrev, T. Heine, T. Frauenheim, and E. Ganz, J. Am. Chem. Soc. **137**, 2757 (2015).
[65] B. Feng *et al.*, Nat. Commun. **8**, 1007 (2017).
[66] L.-M. Yang, I. A. Popov, T. Frauenheim, A. I. Boldyrev, T. Heine, V. Bačić, and E. Ganz, Phys. Chem. Chem. Phys. **17**, 26043 (2015).
[67] L.-M. Yang, I. A. Popov, A. I. Boldyrev, T. Heine, T. Frauenheim, and E. Ganz, Phys. Chem. Chem. Phys. **17**, 17545 (2015).
[68] F. Mouhat and F.-X. Coudert, Phys. Rev. B **90**, 224104 (2014).
[69] S. Zhang, J. Zhou, Q. Wang, X. Chen, Y. Kawazoe, and P. Jena, P. Natl. Acad. Sci. **112**, 2372 (2015).
[70] C. Lee, X. Wei, J. W. Kysar, and J. Hone, Science **321**, 385 (2008).
[71] R. C. Andrew, R. E. Mapasha, A. M. Ukpong, and N. Chetty, Phys. Rev. B **85**, 125428

(2012).
[72] W. Jiang, M. Kang, H. Huang, H. Xu, T. Low, and F. Liu, Phys. Rev. B **99**, 125131 (2019).
[73] W. Kohn, Phys. Rev. Lett. **2**, 393 (1959).
[74] J. Wang, X. Song, X. Shao, B. Gao, Q. Li, and Y. Ma, J. Phys. Chem. C **122**, 27820 (2018).